# Double Correlations in HTSC

**Moshe Dayan**

**Department of Physics, Ben-Gurion University,**

**Beer-Sheva 84105, Israel.**

**E-mail:** mdayan@bgu.ac.il

# Double Correlations in HTSC

**Moshe Dayan**

**Department of Physics, Ben-Gurion University,**

**Beer-Sheva 84105, Israel.**

## Abstract

The Nambu-Gorkov generalized Hartree-Fock theory of superconductivity is further generalized to treat simultaneously two kinds of correlations: the correlations that lead to superconductivity, as well as the correlations that produce the pseudogap in the normal state. The treatment of these simultaneous double correlations is done by enlarging the dimension of the propagator and self-energy matrices, from two to four. The off-diagonal self-energies and the DOS are calculated. The results are compatible with tunneling and PES experiments.

# 1. Introduction

Recently, I have analyzed the pseudogaps of HTSC [1], by means of a matrix propagator that is analogous to the one used by Nambu and Gorkov for the analysis of conventional superconductivity [2,3]. The logic for the choice of this analytical tool is based on the common feature of the two problems. This common feature is electron correlations that lead to phase transition in the superconductive case, and to quasi-phase transition in the normal state of HTSC. In regular superconductivity, Cooper pairs of zero momentum are coherently scattered from state $(\mathbf{k}\uparrow, -\mathbf{k}\downarrow)$ into state $(\mathbf{k'}\uparrow, -\mathbf{k'}\downarrow)$, whereas in HTSC in the normal state, electron-hole pairs of twice the Fermi momentum are coherently scattered from state $(\mathbf{k}\uparrow, \overline{\mathbf{k}}\uparrow)$ into state $(\mathbf{k'}\uparrow, \overline{\mathbf{k'}}\uparrow)$ (Here $\overline{\mathbf{k}} = \mathbf{k} - (\mathbf{k}_{\perp}/|\mathbf{k}_{\perp}|) \cdot 2\mathbf{k}_F$, where $\mathbf{k}_{\perp}$ is the k component normal to the nesting surface, denotes the state which annihilation produced the hole). In superconductivity, the common zero momentum for all Cooper pairs enables maximum amount of coherent scatterings, whereas in HTSC, the coherent scattering is also maximized because of the common momentum of all the electron-hole pairs, which is $(\mathbf{k}_{\perp}/|\mathbf{k}_{\perp}|) \cdot 2\mathbf{k}_F$. This maximization is possible because of the large amount of nesting that is believed to exist in HTSC. In a former paper, Dayan has shown that this nesting causes the divergence of the static electron polarization at $\mathbf{k} = 2\mathbf{k}_F$ [4]. Usually, this kind of irregularity breaks the symmetry of the electron states, and changes them into some condensed states. It has been shown that only in the insolating "mother" materials of HTSC the symmetry-breaking is complete and permanent, whereas in metallic HTSC it is fluctuating [1]. Nevertheless, these fluctuating quasi-condensed states exist, and have profound effects on the properties of HTSC, both in the normal and in the superconductive states. Besides the above correlations, HTSC have correlated Cooper-pairs in the superconductive state. The question is: could these two different kinds of correlations co-exist and if so how could they be treated by the Nambu-Gorkov theory?

The overwhelming experimental evidence suggests that the two kinds of correlations co-exist. Several angle resolved PES experiments indicate clearly the existence of an energy gap in the superconductive state, which becomes a pseudogap in the normal state, instead of disappearing completely at the superconductive transition [5-8]. Tunneling experiments on under-doped samples exhibit energy gaps, at low temperatures, which are far too large to be associated solely to superconductivity [9-11]; they are the (vector) sums of superconductive gaps and normal state pseudogaps.

The correlations of Cooper-pairs are essentially different from the el-hole correlations that are described in Ref. [1]. Such an essential difference is that the superconductive ground state is not the eigenstate of the number operator [3], whereas the ground state of the correlated el-hole system in Ref. [1] is. This essential difference is reflected in the different characteristics of the two systems. This suggests that the Gorkov-Nambu matrix propagator, with only one anomalous (of-diagonal) propagator, is not sufficient to treat the existing problem which has to incorporate two different kinds of symmetry breakings. One has to enlarge the matrix dimension at least by unity. The possible use of 3-dimensional matrices, has not been studied in the present paper, instead, the matrix dimension is increased by two, from two to four. We find this to be more convenient because of the inversion symmetry of the Brillouin zone.

## 2. The field operator, the propagator, and the Hamiltonian.

The components of the field operator in Ref. [1] are $c_{k,s}$ and $c_{\bar{k},s}$. If one wants to include superconductivity, one had to add the Cooper pair matches of the above components in accordance with the Gorkov-Nambu scheme [2,3]. Thus, we define the field operator as

$$\Psi_{k,s} = \begin{pmatrix} c_{k,s} \\ c^+_{-\bar{k},-s} \\ c_{\bar{k},s} \\ c^+_{-k,-s} \end{pmatrix}, \text{ and its conjugate as } \Psi^+_{k,s} = \begin{pmatrix} c^+_{k,s} & c_{-\bar{k},-s} & c^+_{\bar{k},s} & c_{-k,-s} \end{pmatrix}. \qquad (1)$$

The propagator is given by

$$G_s(k,t) = -i \langle \Phi | T \{ \Psi_{k,s}(t) \Psi_{k,s}^+(0) \} | \Phi \rangle \qquad (2)$$

$$= -i \left\langle \Phi \left| T \left\{ \begin{matrix} c_{k,s}(t)c_{k,s}^+(0) & c_{k,s}(t)c_{-\bar{k},-s}(0) & c_{k,s}(t)c_{\bar{k},s}^+(0) & c_{k,s}(t)c_{-k,-s}(0) \\ c_{-\bar{k},-s}^+(t)c_{k,s}^+(0) & c_{-\bar{k},-s}^+(t)c_{-\bar{k},-s}(0) & c_{-\bar{k},-s}^+(t)c_{\bar{k},s}^+(0) & c_{-\bar{k},-s}^+(t)c_{-k,-s}(0) \\ c_{\bar{k},s}^-(t)c_{k,s}^+(0) & c_{\bar{k},s}^-(t)c_{-\bar{k},-s}(0) & c_{\bar{k},s}^-(t)c_{\bar{k},s}^+(0) & c_{\bar{k},s}^-(t)c_{-k,-s}(0) \\ c_{-k,-s}^+(t)c_{k,s}^+(0) & c_{-k,-s}^+(t)c_{-\bar{k},-s}(0) & c_{-k,-s}^+(t)c_{\bar{k},s}^+(0) & c_{-k,-s}^+(t)c_{-k,-s}(0) \end{matrix} \right\} \right| \Phi \right\rangle$$

where $|\Phi>$ is the ground state of the interacting system, and T is the usual time ordering operator. In the following, we will approximate the propagator and its relevant self-energy components by means of the matrix Feynman-Dyson's perturbation theory. For this purpose we use the well-known $\alpha - Dirac$ matrices, which are defined by $\alpha_i = \begin{pmatrix} 0 & \tau_i \\ \tau_i & 0 \end{pmatrix}$, with the Pauli matrices- $\tau_i$ (see definition in Ref. [1]). We also define the four dimensional $\tau_i$ matrices as $\tau_i = \begin{pmatrix} \tau_i & 0 \\ 0 & \tau_i \end{pmatrix}$, and the known Dirac matrix- $\beta = \begin{pmatrix} I & 0 \\ 0 & -I \end{pmatrix}$.

In Ref. [1] we divided the Hamiltonian in two parts, $H^{nes}$ was related only to states in the nested segments of the Fermi surface, including the interactions between them, while $H^{n.n}$ was related to the non-nested segments. Interactions between "mixed" states or between states on planar sections that are not parallel were associated with $H^{n.n}$, and were assumed to result only in the regular renormalization of the eigenstates of the unperturbed Hamiltonian. This simplistic approach is good enough for the first analysis of the quasi-condensed states in the nesting segments. However, it is not adequate for the present problem, since the states in non-nested segments are Cooper-paired, and should participate in the matrix perturbation. Experiments indicate that these states exhibit not only superconductive gaps, but also pseudogaps in the normal state. Moreover, in angle resolve PES the two gaps are apparent only in directions where the non-nested segments are dominants [5-8]. This suggests that the four dimensional perturbation procedure should include all states by the Fermi surface- nested and non-nested. The non-nested states do not participate in the coherent scatterings of the el-hole pairs, but are linked to states in the nested segments of the Fermi surface, through the "mixed" interactions. With this notion the Hamiltonian is $H = H_0 + H_i$, where

$$H_0 = \sum_{k,s} E_k \Psi_{k,s}^+ \beta \Psi_{k,s} \ , \tag{3a}$$

$$H_i = \frac{1}{2} \sum_{k,k',q,s,s'} \{ V[\Psi_{k'-q,s'}^+ \tau_3 \Psi_{k',s'} \Psi_{k+q,s}^+ \tau_3 \Psi_{k,s}]$$

$$+ V_\Lambda [\Psi_{k'-q,s'}^+ \alpha_3 \Psi_{k',s'} \Psi_{k+q,s}^+ \alpha_3 \Psi_{k,s}] + V_\Delta [\Psi_{k'-q,s'}^+ \alpha_1 \Psi_{k',s'} \Psi_{k+q,s}^+ \alpha_1 \Psi_{k,s}] \} . \tag{3b}$$

The sums are on nested and non-nested states over the Brillouin zone. The prefactors in Eqs. (3) are determined in accordance with summation only over half the volume enclosed by the Fermi surface. With this summation scheme every momentum in the first Brillouin zone is presented once in each of the bilinear forms in Eqs. (3).

As we have pointed out in Ref. [1], perturbation theory cannot account for phase transitions unless the symmetry of the condensed phase is already built in the unperturbed Hamiltonian. The problem is that usually the unperturbed Hamiltonian has the eigenstates of the more symmetric phase. The way to get around this difficulty is to add to this Hamiltonian, a one body internal potential Hamiltonian- $H_1$, that is compatible with the condensed phase, such that $H_0^* = H_0 + H_1$, $H_i^* = H_i - H_1$, and $H = H_0 + H_i = H_0^* + H_i^*$. In our case

$$H_1 = \sum_{k,s} \{ \Lambda_0 \Psi_{k,s}^+ \alpha_3 \Psi_{k,s} + \Delta_0 \Psi_{k,s}^+ \alpha_1 \Psi_{k,s} \} \ , \tag{4}$$

where $\Lambda_0$ and $\Delta_0$ are internal potentials that take different values for nested or for non-nested states. Then the equation for the ground state is $H_0^* \Phi_0 = \xi_0 \Phi_0$, with

$$\Phi_0 = \prod_{E_k < 0} [u_k + \frac{v_k}{\sqrt{2}} \Psi_{k,s}^+ \alpha_1 \Psi_{k,s} + \frac{w_k}{\sqrt{2}} \Psi_{k,s}^+ \alpha_3 \Psi_{k,s} + \theta_k (c_{k,s}^+ c_{k,s} c_{-k,-s}^+ c_{-k,-s})] \mid 0 >,$$

(5a)

$$\xi_0 = - \sum_{E_k < 0, s} \sqrt{E_k^2 + \Lambda_0^2 + \Delta_0^2}$$

(5b)

where $\mid 0 >$ is the ground state of $H_0$, and $u_k^2 + v_k^2 + w_k^2 + \theta_k^2 = 1$.

The ground state $\Phi_0$ has the symmetry of the condensed phase, which enables perturbation expansion to establish the condensed phase self-consistently. The validity of the process is also established by the existence of the following anti-commutation relations. The scalar anti-commutation relation is

$$\{\Psi_{k,s}, \Psi_{k',s'}^+\} = 4\delta_{s,s'}\delta_{k,k'} .$$

(6a)

The matrix anti-commutation relation is

$$\{\Psi_{k,s}, \Psi_{k',s'}^+\} = \delta_{s,s'}[\delta_{k,k'}I + M\delta_{k,\bar{k}'}\alpha_3],$$

(6b)

where M is a diagonal matrix. The momentum summation scheme in the present paper implies that $\delta_{k,\bar{k}'} = 0$, so that the right hand side of Eq. (6b) scales with the unit matrix. The following relations are both matrix and scalar

$$\{\Psi_{k,s}, \Psi_{k',s'}\} = \{\Psi_{k,s}^+, \Psi_{k',s'}^+\} = 0 .$$

(6c)

Eq. (3b) suggests three kinds of interaction vertices: $\tau_3$, $\alpha_3$, and $\alpha_1$. The vertex $\tau_3$ needs no special justification since it results in the regular interaction, as the vertex $\tau_3$ in the theory of Gorkov-Nambu [3]. However, the justification of the other two vertices is less trivial. The vertices $\alpha_3$ and $\alpha_1$ scale with the same matrices as the off-diagonal anomalous propagators. It is not hard to see that the assumption of such vertices is the foundation of our theory, since it results in significant off-diagonal self-energies of the Hartree type. Each of these interactions in the Hamiltonian $H_i$ is written as a sum of 16 terms, where each is a product of four c (or $c^+$)-operators. For the interactions with $\alpha_3$ vertices, there are 8 terms which conserve momentum, and 8 terms which violate the momentum conservation by $4k_F$. The momentum conserving terms portray regular interaction between two states on opposite nested sections, which interchange their sections after scattering. An example of such a term is $c^+_{\bar{k}+q,s} c_{k,s} c^+_{k'-q,s} c_{\bar{k}',s}$, when k and k' are on the same nested section, and $\mathbf{q}$ is too small to transfer to the opposite section. The violation of momentum conservation stems from the definition of the vertex $\alpha_3$. Each scattering through this vertex violates momentum conservation by $2k_F$. In the momentum conserving scatterings, two such violations compensate each other, whereas in the non-conserving scatterings the violations add. We immediately see the analogy with electron scatterings in crystalline matter, where the momentum conservation may be violated by a reciprocal lattice vector. The two effects are similar in their violation of momentum conservation, in their connection to symmetry breaking and their resultant condensed phases, and in the internal fields that are associated with them. The obvious difference, though, is that in the discussed case the internal field results from electron correlations, whereas the crystalline field results from the ionic potentials. In 2-dimensional mother insolating materials, $2k_F$ in directions normal to the nested planes is equal to the vectors $\pm(\pi,\pm\pi)$ in the 2-dimensional reciprocal space. These are additional reciprocal vectors that define an additional periodicity with a double period along these directions. This double period is indeed the period of the spin polarization in the anti-ferromagnetic Mott-insulators. For the doped conducting counterparts, experiments show only fluctuations of anti-ferromagnetic domains. The short range fluctuating spin polarizations are of the same period as in the anti-ferromagnetic materials, but we assume that their long range averaged wave-vector is $2k_F$.

The interactions that portray two body scatterings via $\alpha_1$ vertices are derived into 16 terms (of c and $c^+$-operators) for each given k, k', and s. Eight of these terms are interaction terms that conserve the number of particles. The other eight, however, do not conserve the number of particles. An example of such a term is $c_{-k-q,-s} c_{k,s} c_{-k'+q,-s} c_{k',s}$, which annihilates two particles in each vertex of the interaction. Of course, the change of the particles number is averaged to zero when all the other terms are summed up. Nevertheless, when the ground state is the usual metallic one, each term with four annihilation operators (or four production operators) should be unacceptable as a fundamental interaction process. However, here their acceptability is conceivable due to the fact that the superconductive ground state itself is not an eigenstate of the number operator, and our perturbation procedure is based on this ground state. This could be demonstrated by the use of the well-known Bogoliubov-Valatin operators $\gamma_{k,s} = u_k c_{k,s} - v_k c_{-k,-s}^+$, which are the operators of excitations in the superconductive state [3,12]. Suppose that the $\gamma^+$ excitations interact via some scattering potential $V_\gamma$, then when the interaction $V_\gamma(1-2)\gamma_1^+\gamma_2\gamma_3^+\gamma_4$ is expressed in terms of the c's, it contains terms with four annihilation operators (or four creation operators), that do not conserve the number of particles. The above discussion validates in principle all the interactions that are portrayed in $H_i$ of Eq. (3b). The problem that remained to be studied more in detail (in a separate paper) is the nature of the potentials that factorize the various interaction terms.

There are additional components of the matrix propagator of Eq. (2) which should be examined. These are the components that match those of the matrix $\tau_1$. These components portray combinations such as $c_{k,s}(t)c_{-\bar{k},-s}(0)$, which pairs the state (k,s) with the state $(-\bar{k},-s)$, rather than with the state (-k,-s). Could the $\tau_1$ matrix carry an independent self-energy component, and an interaction vertex? For the time being, we could not find any experimental data that supports the addition of such a complication to the theory. However, the idea should be re-examined because there are some considerations (that are not discussed in the present paper) that might favor the introduction of the $\tau_1$ vertex.

The above discussion suggests the applicability of the Feynman-Dyson perturbation theory, with the vector field operators of Eq. (1). The matrix Dyson equation is

$$G^{-1} = G_0^{-1} - \Sigma, \tag{7}$$

where $G_0$ is the diagonal matrix propagator of the unperturbed Hamiltonian, $G_0^{-1}(\omega,k) = [\omega I - \beta E_k + i\delta\omega I]$, and $\Sigma$ is the self-energy matrix. The self-energy has various components but some of them are well known from the regular GHF theory. These known self-energies result in a shift of the Fermi level, and in renormalizations of $\omega$ and $E_k$. For the present analysis, we consider only the $\alpha_1$ and the $\alpha_3$ components of $\Sigma$; $\Sigma = \Lambda\alpha_3 + \Delta\alpha_1$. After multiplying and dividing Eq. (7) by $[\omega I + \beta E_k + \Sigma]$ we get

$$G(\omega,k) = \frac{\omega I + E_k\beta + \Lambda\alpha_3 + \Delta\alpha_1}{\omega^2 - E_k^2 - \Lambda^2 - \Delta^2 + i\delta} \tag{8}$$

## 3. The pseudogap, the superconductive gap, and the density of states.

The pseudogap $\Lambda$ has been calculated in Ref. [1]. The calculation here is not much different in principle. There are some minor differences, which will be pointed out in the following. The basic equation for the pseudogap is obtained in according to the $\alpha_3$ component of the Feynman diagram in Fig. 1a

$$\Lambda_H = -iU_3(\omega = 0)\text{Tr}(\alpha_3^2)\int\frac{d^3pd\nu}{4(2\pi)^4}\frac{\Lambda\exp(i\nu\delta)}{\sqrt{E_p^2 + \Sigma^2}}$$

$$\times \{\frac{1}{\nu - \sqrt{E_p^2 + \Sigma^2 - i\delta}} - \frac{1}{\nu + \sqrt{E_p^2 + \Sigma^2 - i\delta}}\} \ . \tag{9}$$

The extra division by 4 results from the prefactors of the interaction Hamiltonian, and it is in accordance with p-summation over the whole nested parts of the Brillouin zone. Note that, unlike the matrix $\Sigma$ in Eq. (7), in Eq. (9) and hereafter $\Sigma$ is a scalar: $\Sigma = \sqrt{\Lambda^2 + \Delta^2}$. After the $\nu$ integration, and after changing the variable of the p-integration, we get

$$\Lambda_H = -U_3 \int_{-E_m}^{E_m} dE N_0(E) \Lambda (E^2 + \Sigma^2)^{-\frac{1}{2}} \; , \tag{10}$$

where $\pm E_m$ are the upper and lower energy limits for the nested states. Eq. (10) is formally identical with its two dimensional counterpart of Ref. [1], except for the replacement of $\Lambda$ (in the square-root) by $\Sigma$. With the assumption of roughly constant DOS within the integration range, Eq. (10) is integrable, which yields for $\Sigma << E_m$

$$\Lambda_H \cong -2U_3 N_0 \Lambda \ln \frac{2E_m}{\Sigma} \; . \tag{11}$$

Note that the DOS $N_0$ relates only to the nesting sections of the Fermi surface. The calculation of $\Delta_H$ is quite equivalent. A possible difference might stem from different potentials in accordance with the discussion in the former section. Thus, we have

$$\Delta_H \cong -2U_1 N_0 \Delta \ln \frac{2E_m}{\Sigma} \; . \tag{12}$$

Suppose that we assume $U_1 = U_3$, then we get $\frac{\Delta_H}{\Delta} = \frac{\Lambda_H}{\Lambda}$, which leads to $f_\Delta = \frac{\Delta_F}{\Delta} = f_\Lambda = \frac{\Lambda_F}{\Lambda}$. However, we shall see that $\Lambda_F / \Lambda$ and $\Delta_F / \Delta$ have different signs. Therefore, the last relations cannot be correct, and we must conclude that the potentials $U_1$ and $U_3$ are different. The nature of $U_1$ is not clear, because the interaction through the vertex $\alpha_1$ is unusual, and not well studied yet.

Eqs. (11) and (12) may be written as

$$\Sigma = \sqrt{\Lambda^2 + \Delta^2} = 2E_m \exp(\frac{1 - f_\Delta}{2U_1 N_0}) = 2E_m \exp(\frac{1 - f_\Lambda}{2U_3 N_0}) \; . \tag{13}$$

We recall that the potential $U_3$ in Eq. (13) is static, negative, and scatter by $2k_F$. However, the nature of $U_1$ is still not completely clear. We also note that $\Lambda_H$ is obtained by states which are in substantial nesting sections on the Fermi surface. We assume this to be correct also for $\Delta_H$, although this assumption is subject to more analysis of $U_1$. The Fock functions $\Delta_F$ and $\Lambda_F$, though, are not negligible even in the non-nested segments by the Fermi surface.

The calculation of the Hartree parameters that are given by Eqs. (11) and (12) is done for un-doped materials. The introduction of doping to produce conductors and high temperature superconductors, introduces disorder. In ref. [1], we characterized the disorder by introducing into the propagator a single parameter as an imaginary self energy

$$G(\omega, k) = \frac{[\omega + i\gamma s(\omega)]I + E_k \beta + \Lambda \alpha_3 + \Delta \alpha_1}{[\omega + i\gamma s(\omega)]^2 - E_k^2 - \Lambda^2 - \Delta^2}$$

$$= \frac{I[\omega + i\gamma s(\omega)] + \beta E + \alpha_3 \Lambda + \alpha_1 \Lambda}{2B}[(E+B)^{-1} - (E-B)^{-1}] \; , \tag{14}$$

where $B = \sqrt{[\omega - i\gamma s(\omega)]^2 - \Sigma^2}$. We found that the introduction of $\gamma$ did not produce an imaginary part for the pseudogap at zero frequency. It does, however, affect the DOS by producing states in the gap, turning the gaps into pseudogaps in the doped materials. Thus, in the calculation below, we assume the propagator of Eq. (14), rather than the one of Eq. (8).

The Fock integral for $\Lambda_F$ is depicted by the diagram in Fig. 1b, and is given by

$$\Lambda_F(\omega, k) = i \int \frac{dv d^3 p}{(2\pi)^4} \alpha_3 \{G_\Lambda^{nes}(v, p) + G_\Lambda^{n.n}(v, p)\}$$

$$\times \sum_{\xi = \tau_3, \alpha_1, \alpha_3} (\xi \alpha_3 \xi)[\sum_\lambda g_{\lambda, \xi}^2 D_\lambda(\omega - v, k - p) + \exp(iv\delta)V_\xi] \tag{15}$$

where $G_\Lambda^{nes}$ denotes the propagator in the nested segments by the Fermi surface, and $G_\Lambda^{n,n}$ denotes the propagator in the non-nested segments. Accordingly, the p-integral should be performed only over those parts by the Fermi surface in which the propagators are defined. The extra division by 4 results from the prefactors of the interaction Hamiltonian, and it is in accordance with p-summation over the whole Brillouin zone. The potential $V_\xi$ denotes the screened Coulomb interaction, and $g_{\lambda,\xi}$ the screened el-phonon interaction, that correspond to the $\xi$-vertex. The propagator of the $\lambda$-mode phonon is $D_\lambda$. The matrix product $\alpha_3\xi\alpha_3\xi$ is equal to $(-1)^j$, where j = 0 when the matrices commute, and j = 1 when they anti-commute. Therefore, the vertices $\tau_3$ and $\alpha_3$ yield positive multipliers- $\alpha_3\xi\alpha_3\xi$, whereas the vertex $\alpha_1$ yields a negative multiplier.

The integral for $\Delta_F$ is obtained similarly by

$$\Delta_F(\omega,k) = i\int\frac{dv d^3p}{(2\pi)^4}\alpha_1\{G_\Delta^{nes}(v,p) + G_\Delta^{n,n}(v,p)\}$$

$$\times \sum_{\xi=\tau_3,\alpha_1,\alpha_3}(\xi\alpha_1\xi)[\sum_\lambda g_{\lambda,\xi}^2 D_\lambda(\omega-v,k-p) + \exp(iv\delta)V_\xi]. \tag{16}$$

Here, too, the p-integration of the first term of Eq. (16) is over the nested parts, and of the second term is over the non-nested parts.

Eqs. (15) and (16) may be derived further by following the same steps as in the regular generalized Hartree-Fock theory (GHF). These steps include the "contour bending" in the $v$ plane, and the momentum integration [1, 3]. Then we get

$$\Lambda_F(\omega,k) = N_0^{nes}\int_0^\infty dv\,Re\,\frac{\Lambda^{nes}}{B^{nes}(v)}\{-\alpha^2 F_k^{nes}[(v+\omega+\Omega-i\delta)^{-1} + (v-\omega+\Omega-i\delta)^{-1}] + \overline{V}_k^{nes}\}$$

$$+ N_0^{n.n} \int_0^\infty d\nu \, \text{Re} \, \frac{\Lambda^{n.n}}{B^{n.n}(\nu)} \{-\alpha^2 F_k^{n.n}[(\nu + \omega + \Omega - i\delta)^{-1} + (\nu - \omega + \Omega - i\delta)^{-1}] + \overline{V}_k^{n.n}\}$$

$$(17)$$

$$\Delta_F(\omega, k) = N_0^{nes} \int_0^\infty d\nu \, \text{Re} \, \frac{\Delta^{nes}}{B^{nes}(\nu)} \{\alpha^2 F_k^{nes}[(\nu + \omega + \Omega - i\delta)^{-1} + (\nu - \omega + \Omega - i\delta)^{-1}] - \overline{V}_k^{nes}\}$$

$$+ N_0^{n.n} \int_0^\infty d\nu \, \text{Re} \, \frac{\Delta^{n.n}}{B^{n.n}(\nu)} \{\alpha^2 F_k^{n.n}[(\nu + \omega + \Omega - i\delta)^{-1} + (\nu - \omega + \Omega - i\delta)^{-1}] - \overline{V}_k^{n.n}\} .$$

$$(18)$$

In Eqs. (17) and (18), $N_0^{nes} = N_0$ - is the band DOS of the nested states, $N_0^{n.n}$ is the band DOS of the non-nested states,

$$B^{nes}(\nu) = \sqrt{[\nu + i\gamma s(\nu)]^2 - (\Sigma^{nes})^2} \ , \tag{19a}$$

$$B^{n.n}(\nu) = \sqrt{[\nu + i\gamma s(\nu)]^2 - (\Sigma^{n.n})^2} \ , \tag{19b}$$

and $\alpha^2 F_k^{nes,n.n}$, and $\overline{V}_k^{nes,n.n}$ are integral averaged values, which are defined by Eqs. (15) through (18). The quantities $\Lambda^{nes}$ and $\Delta^{nes}$ are the self energies that correspond to the nesting sections of the Fermi surface, whereas $\Lambda^{n.n}$ and $\Delta^{n.n}$ are the self-energies that correspond to the non-nesting segments.

Note that Eqs. (15) through (18) are valid regardless whether k is within a nested segment or non-nested segment by the Fermi surface. This is contrast to the Hartree integrals, which are assumed to be small for k within a non-nested segment by the Fermi surface. Note also that the Fock integrals for k's in non-nested segments gain substantial contributions from integration on nested states too. Actually, Fock integrals for k's in nested or non-nested segments differ only by the averaged interactions $\alpha^2 F_k$ and $\overline{V}_k$. One should also notice that the signs of $\alpha^2 F_k$ and $\overline{V}_k$, in the integrals of $\Lambda_F$ are reversed relative to the signs in the integrals of $\Delta_F$. This means that for $\Lambda$ and $\Delta$ of the same sign, one gets $\Lambda_F$ and $\Delta_F$ of different signs. This should make $\Delta$ smaller than $\Lambda$, for comparable interaction potentials. This relation, which agrees with experiment, is reflected from the nested sections to the non-nested sections.

Experiments that probe the electronic DOS are of essential importance for revealing the mysteries of HTSC. Therefore, we wish to calculate the DOS that results from the theory. The general equation for the total DOS is

$$N(v) = \frac{-s(v)}{\pi} \operatorname{Im} \int \frac{d^3p}{(2\pi)^3} G_1(v, \mathbf{p}) .$$

(20)

The recent achievements of angle resolved photo-emission spectroscopy (ERPES) experiments have made them an important tool for the analysis of HTSC. Eq. (20) is not adequate for comparison with ERPES data, since it provides the total DOS, after angular integration. Besides, for the present analysis, it is essential to provide "directional" DOS in order to discriminate between the DOS of the nested segments of the Fermi surface, to the non-nested ones. Accordingly, we write

$$N_{\delta p}^{nes}(v) = \frac{-s(v)}{\pi} \operatorname{Im} \int_{\delta p}^{nes} \frac{d^3p}{(2\pi)^3} G_1^{nes}(v, \mathbf{p}) ,$$

(21)

$$N_{\delta p}^{n.n}(v) = \frac{-s(v)}{\pi} \operatorname{Im} \int_{\delta p}^{n.n} \frac{d^3p}{(2\pi)^3} G_1^{n.n}(v, \mathbf{p}),$$

(22)

where the superscripts and integration limits of "nes" and "n.n" denote the relevant sector on the Fermi surface, and $\delta p$ denotes the angular section over which the states are to be integrated. We are interested in the symmetry directions that are usually probed in ARPES, and in energies very close to the Fermi level. Typically, for a 2-dimentional HTSC, the most interesting directions are $(0, \pm\pi)$, $(\pm\pi, 0)$, and $\pm(\pi, \pm\pi)$. According to the most basic band theory of HTSC, the nesting states are centered on the directions $\pm(\pi, \pm\pi)$, whereas the non-nested states are centered on the directions $(0, \pm\pi)$ and $(\pm\pi, 0)$.

The integrations of Eqs. (21) and (22) are performed in the standard way of Ref. [1]. Here the results scale with $\delta p / (\delta p)_{\text{total}}^{\text{nes,n.n}}$, where $(\delta p)_{\text{total}}^{\text{nes,n.n}}$ is the total integrated angle for the corresponding category (nested or non-nested). Thus, we have

$$N_{\delta p}^{\text{nes}}(\nu) = N_0^{\text{nes}} \frac{\delta p}{(\delta p)_{\text{total}}^{\text{nes}}} \, \text{Re} \{ \frac{|\nu| + i\gamma}{\sqrt{[\nu + i\gamma s(\nu)]^2 - (\Sigma^{\text{nes}})^2}} \}, \qquad (23)$$

$$N_{\delta p}^{\text{n.n}}(\nu) = N_0^{\text{n.n}} \frac{\delta p}{(\delta p)_{\text{total}}^{\text{n.n}}} \, \text{Re} \{ \frac{|\nu| + i\gamma}{\sqrt{[\nu + i\gamma s(\nu)]^2 - (\Sigma^{\text{n.n}})^2}} \} \, . \qquad (24)$$

We recall that for the non-nested part of the Fermi surface only the Fock integrals are significant, $(\Sigma^{\text{n.n}})^2 = (\Lambda_F^{\text{n.n}})^2 + (\Delta_F^{\text{n.n}})^2$. For the nested part, though, the Hartree self-energies $\Lambda_H$ and $\Delta_H$ are significantly large, which gives $(\Sigma^{\text{nes}})^2 = (\Lambda_H + \Lambda_F^{\text{nes}})^2 + (\Delta_H + \Delta_F^{\text{nes}})^2$. Consequently, $\Sigma^{\text{nes}}$ is much larger than $\Sigma^{\text{n.n}}$.

## 4.    Concluding remarks

In a former analysis we found that a substantial nesting area cannot coexist with a metallic Fermi surface [4], since this causes divergences of some electronic polarizations (at zero energy) and the instability of the electronic symmetry [13]. Since band theory suggests that oxide HTSC should be metals with substantial nesting, we have concluded that a symmetry breakdown is certain, and employed the Nambu-Gorkov theory to analyze this new electronic symmetry [1]. This recent work provides an adequate analysis for the pseudogaps in HTSC. In the present analysis we have enlarged the dimension of the propagator matrix to enable the analysis of two kinds of correlations, which act simultaneously and result in two order parameters: the superconductive gap, and the normal state pseudogap. The crucial proposal of the interaction vertex $\tau_1 (\alpha_3$ in the present notation) which violets momentum conservation by $2k_F$, has been extended to include the $\alpha_1$ vertex, which violets the conservation of particle number. These vertices yield Hartree self-energies, which are of the order of 200meV, because they scale with the large electronic energy scale $E_m$. These large Hartree self-energies are assumed to prevail only in the nested segments by the Fermi surface. They enhance the Fock self-energies in their segments and in the non-nested segments as well.

The present analysis is in a preliminary stage, as it presents only zero temperature self energies and DOS. Therefore, at present, the experiments that it should be confronted with are those which probe low temperature electronic DOS. The most prominent among these are electron tunneling and photo-electron spectroscopy. Data from these experiments should be compared with Eqs. (23) and (24). I am not aware of substantial and reliable tunneling experiments that present directional tunneling data, namely experiments where tunneling results, taken along different crystalline (in plane) directions, are compared. Usually, tunneling data are directionally averaged as they are obtained on polycrystalline samples, or on crystalline Bi2212 surfaces normal to the Bi-O layers [9-11,14-16]. These provide angular averages in which the nested part is weakly weighted at low energy $(\nu \ll \Sigma^{nes})$, due to the large absolute value of its denominator. Consequently, low energy tunneling data is mostly $N_{\delta p}^{n,n}(\nu \ll \Sigma^{nes})$. Thus, the "gap-edge" structure that is usually presented comes mostly from the non-nested states, and is mostly the signature of $\Sigma^{n,n} = \sqrt{(\Lambda_F^{n,n})^2 + (\Delta_F^{n,n})^2}$ .

Since the early tunneling measurements on HTSC, the investigators realized that tunneling results are incomparable with the BCS equation for the DOS. There are several features that deviate from the BCS theory, or its strong coupling counterpart, but here we are referring only to the smearing of the gap-edge peak, and the final DOS at zero bias. The investigators found that the data could be fitted to the empirical DOS function [14-16]

$$N(\nu) = N(\infty) \operatorname{Re} \left\{ \frac{\nu - i\Gamma}{\sqrt{(\nu - i\Gamma)^2 - \Delta^2}} \right\} . \qquad (25)$$

The parameter $\Gamma$ is usually presented as an excitation lifetime parameter. Note that the sign of $i\Gamma$ is reversed relative to the sign of $i\gamma s(\nu)$ in Eqs. (23) and (24), and only the latter is compatible with the analytical properties of G. The effect of the reversed sign is neutralized, only if the square-root is calculated with a branch cut along the negative half of the $\nu$ real axis. Besides, reasonably good fits to Eq. (25) have usually been found for positive energy, but the equation is incorrect for negative energy (unless $\nu$ is replaced by its absolute value). After these modifications, Eq. (25), as well as Eq. (24), yields final DOS at zero energy, in addition to good fits to the broadening of the gap-edge. The good fits that have been found to Eq. (25) confirm the adequacy of Eq. (24). However, the differences between the equations are essential. Eq. (24) has been derived from basic principles, whereas Eq. (25) is phenomenological. The "gap" obtained from the fit to the measured data should be interpreted as $\Sigma^{n,n} = \sqrt{(\Lambda_F^{n,n})^2 + (\Delta_F^{n,n})^2}$ , rather than $\Delta$ , and the broadening parameter $\gamma$ (or $\Gamma$ ) is not the excitation lifetime, but the smearing of the k-states due to elastic scatterings by impurities. The excitation lifetime is also presented in Eqs. (23) and (24) as $\text{Im}\,\Sigma_F^{nes}$ and $\text{Im}\,\Sigma_F^{n,n}$ .

Another confirmation to Eqs. (23) and (24) from tunneling measurements is the large measured ratio $2\Delta / kT_c$ . The more under-doped is the measured sample, the larger is the ratio [9-11,14-16]. These large ratios are compatible with Eq. (24), and with $(\Sigma^{n,n})^2 = (\Lambda_F^{n,n})^2 + (\Delta_F^{n,n})^2$ .

A relevant question is regarding the experimental confirmation of the smeared peak that is predicted by Eq. (23). This peak is actually apparent in tunneling data. It is quite clear in the data of Miyakawa et al. [10], where it is referred to as a "Hump" (around 150meV in inset in Fig. 1, for example). It also appears in other tunneling data [9,15]. However, the appearance of this peak is sometimes obscured due to some wide broadening. This broadening comes from the large imaginary part of the denominator in the right hand side of Eq. (23), as one can see by carefully studying Eqs. (17), (18), and (23).

Another experimental tool, which has become very important for investigating HTSC, is PES- especially its angular resolved version. Before proceeding with the comparison to this experiment, we must stress its limitations. The absorption depth of the photons used in these experiments (20-50eV) is two orders of magnitude larger than the inelastic mean-free-path of the photoelectrons. The latter is considered to be a "universal function" for all materials, and its value for the discussed energy range is only a few A°. Consequently, only a very small fraction of the emitted electrons are primary photoelectrons. The most are secondary electrons which underwent some inelastic scattering. The larger is the binding energy, the larger is the fraction of the secondary electrons. The counter argument to this criticism is that the inelastically scattered photoelectrons, being scattered by electron excitations, appear only at large binding energies of the spectrum. This argument underestimates the electron-phonon scatterings. We estimate that the ratio between the latter to the total number of inelastic scatterings is large enough to suggest that any emitted electron with binding energy larger than the leading frontal step is strongly suspected of being a secondary electron [17]. Consequently, the analysis of PES data in terms of the electronic spectral function is meaningless (or at least should be interpreted very carefully), for binding energies larger that the first few tens of meV. However, for the analysis of the leading edge the tool is adequate and effective, if interpreted carefully. This is especially true for the high resolution ARPES, where gaps and pseudogaps have been measured versus the main symmetry directions of the Brillouin zone [5-8].

The contribution of ARFES to the understanding of HTSC is mainly in two categories: The first is the establishment of the pseudogaps in the normal state. The second is the discovery that both the superconducting gap and the pseudogap are detectable only close to the direction $(0, \pi)$, and its equivalent. No gaps have been observed in the direction $(\pi, \pi)$, but given the limitations of ARFES, this conclusion is applicable only to small binding energies. The angular dependence of ARPES data have been considered as a support to theories that proposed gaps with the $d_{x^2 - y^2}$ symmetry. Our results are not so specific in terms of the angular dependence of the gap function, but they are given in terms of two different directional categories: the nesting direction ("nes"), and the non-nesting direction ("n.n"). All band structure works suggest that the first is associated with the $(\pi, \pi)$ direction, while the second is associated with the $(0, \pi)$ direction. First we notice that both Eqs. (23) and (24) suggest states down to zero energy, a fact which has been confirmed by tunneling experiments, too. The data observed by ARPES in the $(0, \pi)$ direction do not contradict this feature, it merely shows a high leading step (smeared by $\gamma$) at $\Sigma^{n,n} = \sqrt{(\Lambda_F^{n,n})^2 + (\Delta_F^{n,n})^2}$, as predicted by Eq. (24). The leading edge at zero energy carries much smaller intensity, and is overshadowed by the gap-edge. However, Eq. (23) suggests that the "gap-edge" peak in the direction $(\pi, \pi)$ is expected to appear at much higher energy ($\Sigma^{nes}$ is estimated to be around 200meV). At this energy, the limitations of the experimental tool become effective, and this leading gap-edge is covered and concealed by secondary electrons. Besides, this peak is harder to observe due to the extra broadening that discussed before.

We conclude by asserting that our theory yields both superconductive gap, and pseudogap in the normal state. The method proposed is capable of obtaining large values for these order parameters, despite the low DOS at the Fermi level. This is due to the large Hartree off-diagonal self-energies with their large electronic prefactor $E_m$. The last discussion demonstrates agreement with tunneling and ARPES measurements.

Fig. 1

The two diagrams that depict the off-diagonal self-energy. The matrix component of the self-energy is determined by the component of the propagator. A. The Hartree diagram. B. The Fock diagram. Full lines denote electron propagators, and dashed lines denote screened Coulomb plus el-phonon-el interaction.

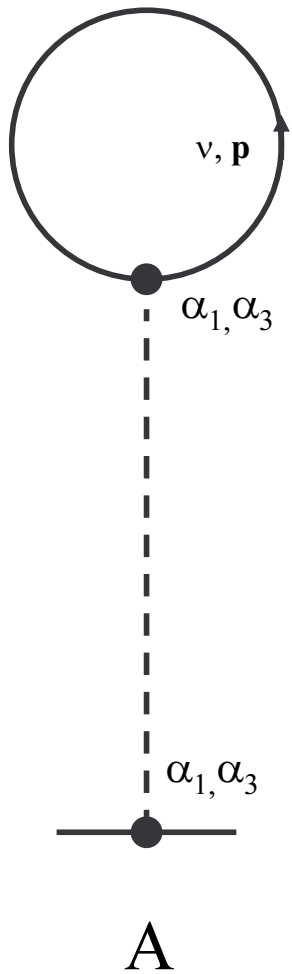

A

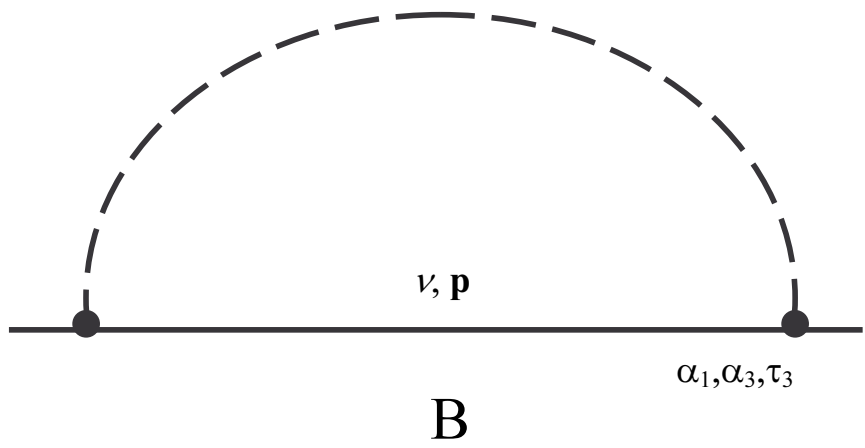

B